\documentclass[conference]{IEEEtran}
\IEEEoverridecommandlockouts

\usepackage{amsmath,amssymb,amsfonts}
\usepackage{algorithm}
\usepackage{algorithmic}
\usepackage{graphicx}
\graphicspath{ {Figures/} }
\usepackage{stfloats}
\usepackage{multicol}
\usepackage{multirow}
\usepackage{tablefootnote}
\usepackage[flushleft]{threeparttable}
\usepackage{color, colortbl}
\usepackage{lipsum}

\usepackage{hyperref}

\usepackage{balance}

\usepackage{cite}

\definecolor{Gray}{gray}{0.9}
\definecolor{Highlight}{rgb}{0.99, 0.99, 0.9}

\usepackage{textcomp}
\usepackage{xcolor}

\def\BibTeX{{\rm B\kern-.05em{\sc i\kern-.025em b}\kern-.08em
    T\kern-.1667em\lower.7ex\hbox{E}\kern-.125emX}}
    
\begin{document}

\title{A Unified Hardware Accelerator for Fast Fourier Transform and Number Theoretic Transform
\thanks{\textcopyright $\,$ 2025 IEEE. Personal use of this material is permitted. Permission from IEEE must be obtained for all other uses, in any current or future media, including reprinting/republishing this material for advertising or promotional purposes, creating new collective works, for resale or redistribution to servers or lists, or reuse of any copyrighted component of this work in other works.}
\thanks{A revised version of this paper was published in the proceedings of the 2025 IEEE International Conference on Acoustics, Speech and Signal Processing (ICASSP) - DOI: \href{https://dx.doi.org/10.1109/ICASSP49660.2025.10889132}{10.1109/ICASSP49660.2025.10889132}}
}

\author{
\IEEEauthorblockN{Rishabh Shrivastava, Chaitanya Prasad Ratnala, Durga Manasa Puli and Utsav Banerjee}
\IEEEauthorblockA{Electronic Systems Engineering, Indian Institute of Science, Bengaluru, India \\
Email: rishabhs@iisc.ac.in, chaitanyar@alum.iisc.ac.in, durgamanasa@alum.iisc.ac.in, utsav@iisc.ac.in}
}

\maketitle

\begin{abstract}
The Number Theoretic Transform (NTT) is an indispensable tool for computing efficient polynomial multiplications in post-quantum lattice-based cryptography.
It has strong resemblance with the Fast Fourier Transform (FFT), which is the most widely used algorithm in digital signal processing.
In this work, we demonstrate a unified hardware accelerator supporting both 512-point complex FFT as well as 256-point NTT for the recently standardized NIST post-quantum key encapsulation and digital signature algorithms ML-KEM and ML-DSA respectively.
Our proposed architecture effectively utilizes the arithmetic circuitry required for complex FFT, and the only additional circuits required are for modular reduction along with modifications in the control logic.
Our implementation achieves performance comparable to state-of-the-art ML-KEM / ML-DSA NTT accelerators on FPGA, thus demonstrating how an FFT accelerator can be augmented to support NTT and the unified hardware can be used for both digital signal processing and post-quantum lattice-based cryptography applications. 
\end{abstract}


\section{Introduction}
\label{sec:introduction}

Modern public key cryptography, based on the intractability of integer factorization and discrete logarithms, is vulnerable to future quantum adversaries capable of realizing Shor's algorithm \cite{shor_quantum_1997} in large-scale fault-tolerant quantum computers.
Therefore, recent advances in quantum computing technologies have motivated the design and implementation of new quantum-secure cryptographic algorithms, also known as \textit{post-quantum cryptography}.
The U.S. National Institute of Standards and Technology (NIST) has been driving the standardization of post-quantum cryptography (PQC) since 2016 \cite{nist_pqc0_2016}.
After several rounds of theoretical analysis, security evaluation and optimized implementation \cite{nist_pqc1_2019, nist_pqc2_2020, nist_pqc3_2020}, NIST has selected CRYSTALS-Kyber \cite{crystals_kyber_2021} and CRYSTALS-Dilithium \cite{crystals_dilithium_2021} as its primary recommendations for quantum-secure key encapsulation mechanism (KEM) and digital signature algorithm (DSA) respectively.
These algorithms have also been included in the recently announced NIST PQC standards ML-KEM \cite{nist_mlkem_2023} and ML-DSA \cite{nist_mldsa_2023} respectively.

Both ML-KEM (Kyber) and ML-DSA (Dilithium) are lattice-based cryptographic algorithms based on the hardness of the Module-LWE (Learning With Errors) problem \cite{peikert_decade_2016}.
They involve the computation of several polynomial multiplications which can be efficiently implemented using the \textit{Number Theoretic Transform} (NTT). The NTT is a generalization of the widely used \textit{Fast Fourier Transform} (FFT) \cite{cormen_algo_2009}. All arithmetic in NTT is performed with integers in a finite field, while FFT requires the use of complex numbers. 

Previous work has presented hardware implementations of NTT using custom accelerators as well as re-purposing existing cryptographic co-processors \cite{banerjee_sapphire_2019, banerjee_iscas_2020, bos_usenix_2022, zhang_iscas_2021, azarderakhsh_arith_2021, land_cardis_2022, gupta_tcas1_2023, aikata_tcas1_2023, li_tvlsi_2024, mandal_vlsid_2024}.
However, there has been little work in exploring how FFT hardware can be enhanced to implement NTT. Recently, \cite{thanawala_fft_2023} has proposed a high-level synthesis framework for polynomial multiplication using pipelined FFT architectures.

In this work, we present the FPGA-based implementation of a unified hardware accelerator supporting fixed-point complex FFT along with NTT for ML-KEM and ML-DSA parameters.
We leverage the similarity between FFT and NTT butterfly operations to demonstrate that a traditional fixed-point complex FFT hardware accelerator can be augmented to achieve ML-KEM and ML-DSA NTT performance comparable to state-of-the-art custom NTT accelerators \textcolor{black}{at the cost of $\approx 62$\% more LUTs, $\approx 26$\% more FFs, and no additional DSPs and BRAMs compared to the baseline FFT. }
Our proposed unified architecture will enable digital communication systems\textcolor{black}{, e.g., wireless sensor nodes, which use FFT accelerators for data processing, to efficiently upgrade their hardware and also support PQC NTT for quantum-secure cryptographic protocols such as key encapsulation and digital signature-based authentication. }

\section{Background}
\label{sec:background}

The FFT is an efficient algorithm used to compute the discrete Fourier Transform of sequences in digital signal processing, e.g., conversion from time domain to frequency domain. Given an $N$-length sequence of complex numbers $\{ \, x_0, x_1, \cdots , x_{N-1} \, \}$, its FFT is given by another $N$-length sequence of complex numbers $\{ \, X_0, X_1, \cdots , X_{N-1} \, \}$ where
\[
X_k = \sum_{m=0}^{N-1} \, x_m \, \omega_N^{mk} \,\,\, \text{for} \,\,\, k \in \{ \, 0, 1, \cdots , N-1 \, \}
\]
Here, $\omega_N = exp(-2 \pi j/N)$ is the $N$-th complex root of unity ($j = \sqrt{-1}$), and its powers $\omega_N^{mk}$ are known as twiddle factors.
Similarly, given a polynomial $a(x) \in \mathbb{Z}_q[x]/(x^N+1)$ with coefficients $a(x) = (a_0, a_1, \cdots, a_{n-1})$, its NTT representation is given by $\hat{a}(x) = (\hat{a}_0, \hat{a}_1, \cdots, \hat{a}_{N-1})$ where
\[
\hat{a}_k = \sum_{m=0}^{N-1} \, a_{m} \, \zeta_N^{mk} \, \text{mod} \, q \,\,\, \text{for} \,\,\, k \in \{ \, 0, 1, \cdots , N-1 \, \}
\]
Here, $\zeta_N$ is the $N$-th primitive root of unity in the ring $\mathbb{Z}_q$, that is, $\zeta_N^N = 1 \, \text{mod} \, q$ and $\zeta_N^k \ne 1 \, \text{mod} \, q$ for $k \ne N$, and its powers $\zeta_N^{mk}$ are also known as twiddle factors.
The modulus $q$ must be chosen to be a prime such that $q \equiv 1 \, \text{mod} \, N$ in order to have elements of order $N$.
Furthermore, to support negative-wrapped convolution \cite{banerjee_sapphire_2019} for polynomial multiplication, the modulus must also satisfy $q \equiv 1 \, \text{mod} \, 2N$ so that both the $N$-th and $2N$-th primitive roots of unity modulo $q$ exist.

Both FFT and NTT algorithms exploit the cyclic properties of $\omega_N$ and $\zeta_N$ respectively to efficiently compute the transform using a series of \textit{butterfly} operations.
There are two butterfly configurations -- Cooley-Tukey (or Decimation-in-Time) and Gentleman-Sande (or Decimation-in-Frequency). In this work, we implement the former which computes $(a \pm \omega \times b)$ and $(a \pm \zeta \times b) \; \text{mod} \; q$ for FFT and NTT respectively, where $a$ and $b$ are inputs to the butterfly, and $\omega$ and $\zeta$ denote the twiddle factors. The overall FFT or NTT computation requires $\frac{N}{2} \text{log}_2 N$ butterflies equally distributed across $\text{log}_2 N$ stages.

For ML-DSA (Dilithium), the NTT parameters are $N = 256$ and $q = 8380417 = 2^{23} - 2^{13} + 1$. Therefore, it involves NTT computation with 256 23-bit polynomial coefficients and $\zeta_N = 3073009$. For ML-KEM (Kyber), the NTT parameters are $N = 256$ and $q = 3329 = 13 \cdot 2^8 + 1$. However, $q \not\equiv 1 \, \text{mod} \, 2N$ in this case, so the input polynomial is interpreted as 128 pairs of coefficients. Therefore, it involves NTT computation with 128 pairs of 12-bit polynomial coefficients and $\zeta_N = 17$. Further mathematical details of Dilithium and Kyber NTTs are available in \cite{crystals_dilithium_2021, nist_mldsa_2023} and \cite{crystals_kyber_2021, nist_mlkem_2023} respectively.

\section{Hardware Architecture}
\label{sec:architecture}

\begin{figure}[!t]
\centering
\includegraphics[width=3.4in]{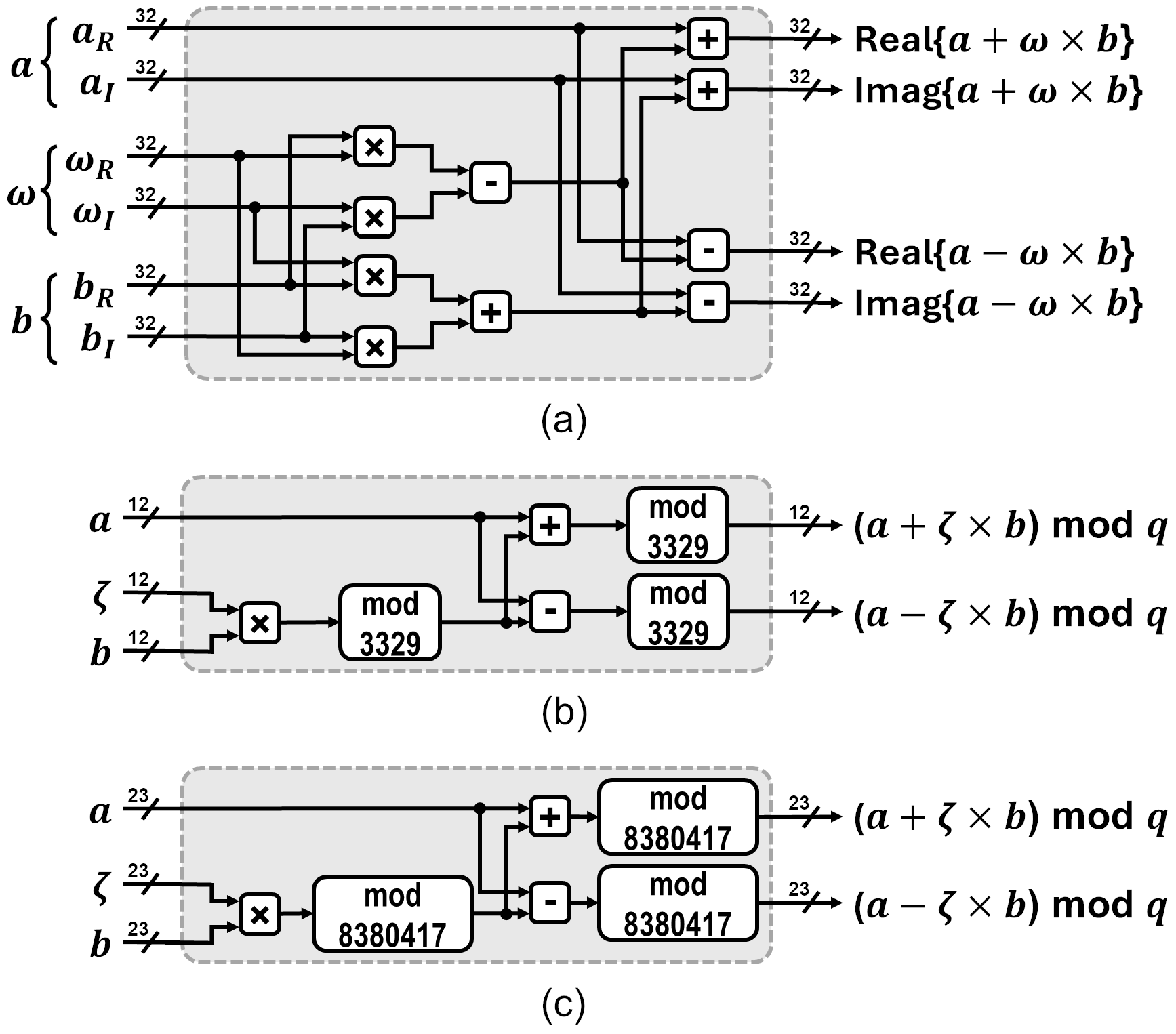}
\caption{Overview of butterfly computations in (a) FFT, (b) ML-KEM (CRYSTALS-Kyber) NTT and (c) ML-DSA (CRYSTALS-Dilithium) NTT.}
\label{fig:butterfly}
\end{figure}

To implement our unified accelerator, we begin with the FFT hardware as baseline and then include additional logic to add support for ML-KEM and ML-DSA NTTs.
In this work, we consider 512-point complex FFT with signed inputs in 32-bit fixed-point Q16.15 format, that is, each of real and imaginary parts has 1 sign bit, 16 integer bits and 15 fractional bits. The complex twiddle factors $\omega$ (where $| \, \omega \, | \le 1$) are represented in 32-bit fixed-point Q1.30 format, that is each of real and imaginary parts has 1 sign bit, 1 integer bit and 30 fractional bits.
All negative quantities are assumed to be stored in standard two's complement format.
Now, the complex FFT butterfly expression can be expanded as:
\begin{align*}
a \pm \omega \times b &= (a_R + ja_I) \pm (\omega_R + j\omega_I) \times (b_R + jb_I) \\
&= (a_R \pm \omega_R b_R \mp \omega_I b_I) + j (a_I \pm \omega_R b_I \pm \omega_I b_R)
\end{align*}
where $a_R$, $b_R$ and $\omega_R$ are the real parts of $a$, $b$ and $\omega$ respectively, and $a_I$, $b_I$ and $\omega_I$ are their complex parts respectively. This requires four 32-bit fixed-point multiplications and several additions / subtractions, as shown in Fig. \ref{fig:butterfly}a. 
The complete 512-point FFT computation involves 9 stages with 256 butterflies per stage. Due the cyclic properties of twiddle factors, 512 complex signed 32-bit twiddle factors are required for the 512-point FFT computation.

\begin{figure}[!t]
\centering
\includegraphics[width=3.4in]{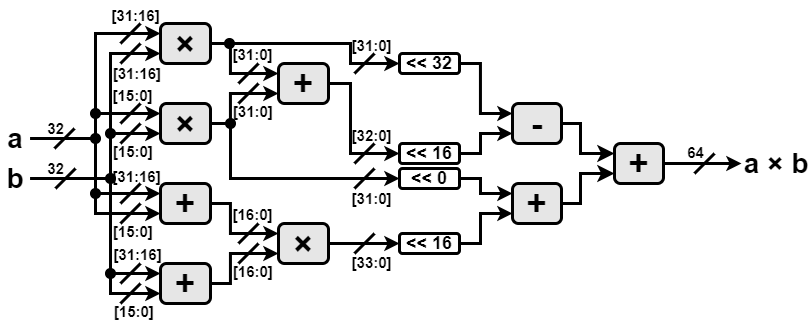}
\caption{Design of 32-bit Karatsuba multiplier using 16-bit multipliers.}
\label{fig:karatsuba_multiplier}
\end{figure}

\begin{figure}[!t]
\centering
\includegraphics[width=2.7in]{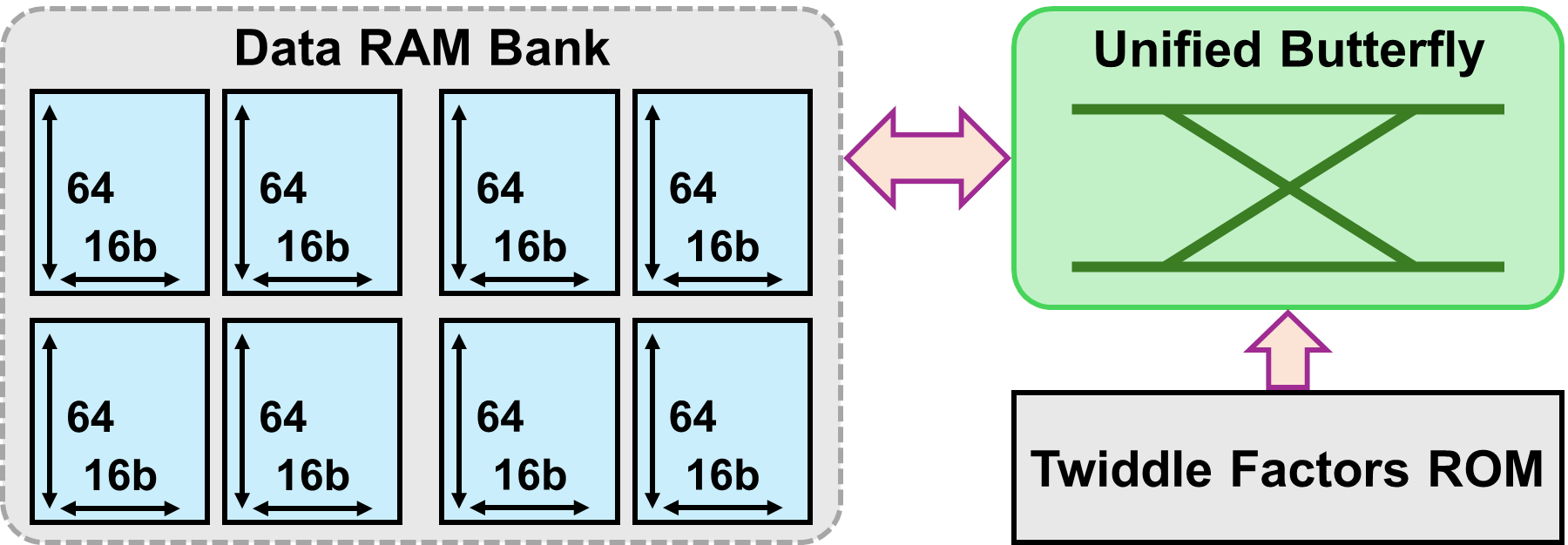}
\caption{Top-level architecture of the proposed unified accelerator.}
\label{fig:top_arch}
\end{figure}

Next, we analyze the computational requirements of ML-KEM and ML-DSA NTT butterfly operations.
For ML-KEM butterfly $(a \pm \zeta \times b) \; \text{mod} \; q$, all inputs and outputs are unsigned 12-bit integers with arithmetic modulo $q = 3329$, as shown in Fig. \ref{fig:butterfly}b. Each ML-KEM NTT involves 7 stages with 64 pairs of identical butterflies per stage. 
Total 128 unsigned 12-bit integer twiddle factors are required for the NTT.
For ML-DSA butterfly $(a \pm \zeta \times b) \; \text{mod} \; q$, all inputs and outputs are unsigned 23-bit integers with arithmetic modulo $q = 8380417$, as shown in Fig. \ref{fig:butterfly}c. Each ML-DSA NTT involves 8 stages with 128 butterflies per stage.
Total 256 unsigned 23-bit integer twiddle factors are required for the NTT.
We observe that ML-KEM NTT requires 12-bit multiplications while ML-DSA NTT requires 23-bit multiplications.
Therefore, we effectively utilize the arithmetic circuits in the 32-bit fixed-point complex FFT by implementing its 32-bit multipliers in Karatsuba configuration \cite{karatsuba_mul_1962} with several 16-bit multipliers, as shown in Fig. \ref{fig:karatsuba_multiplier}.
Similarly, 32-bit additions / subtractions are split into pairs of 16-bit additions / subtractions with appropriate carry / borrow propagation logic. Butterfly inputs in ML-KEM and ML-DSA NTT are zero-padded to 16-bit and 32-bit respectively for the arithmetic operations. This approach requires no additional arithmetic modules except the modular reduction circuits specific to NTT. While Barrett reduction \cite{barrett_red_1986} is used to reduce the multiplication outputs, simple conditional subtraction / addition can be used to reduce the addition / subtraction outputs.

\begin{figure*}[!t]
\centering
\includegraphics[width=7.0in]{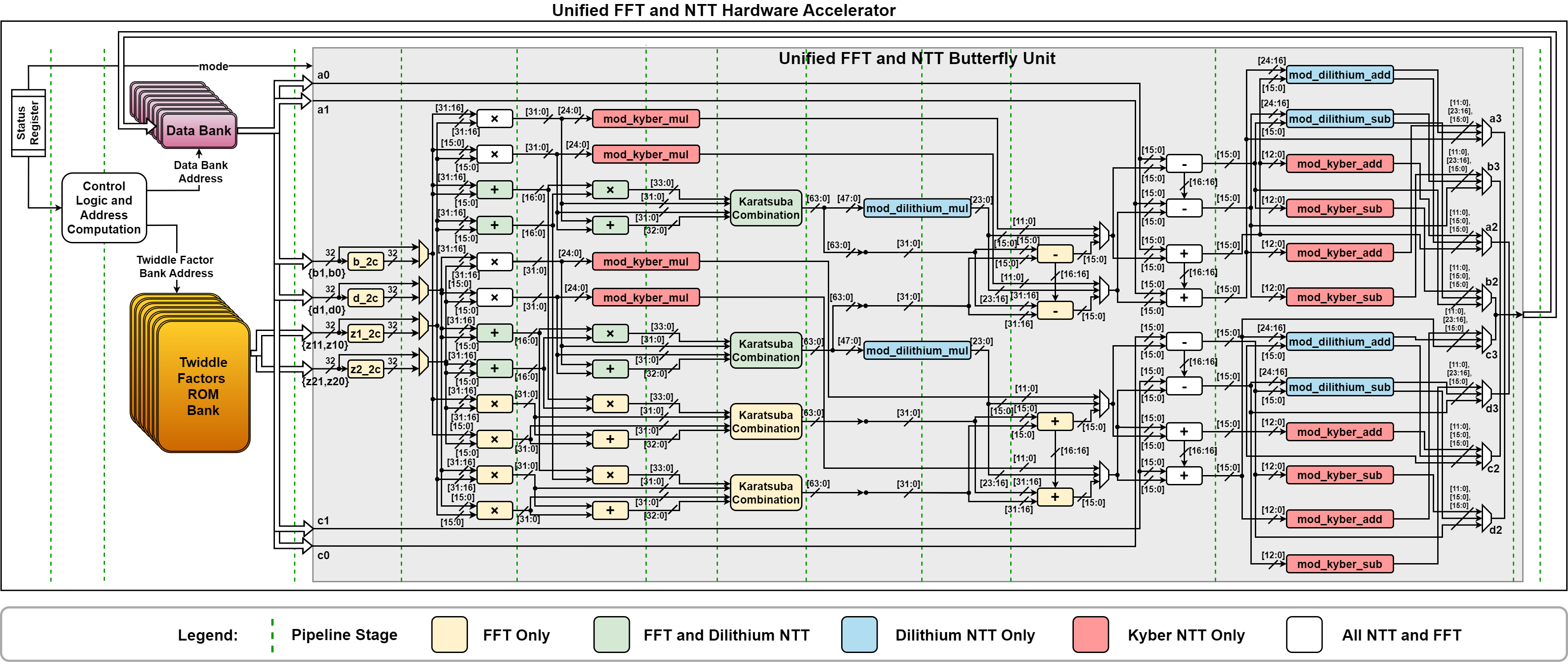}
\caption{Detailed architecture of the proposed unified butterfly unit and memory organization for FFT and ML-KEM (Kyber) + ML-DSA (Dilithium) NTT. The pipeline stages are shown as green dashed lines. The sub-modules of the unified butterfly unit used for FFT only, FFT + ML-DSA NTT, ML-DSA NTT only, ML-KEM NTT only and all NTT + FFT are shown in yellow, green, blue, red and white respectively.}
\label{fig:detailed_arch}
\end{figure*}

Finally, we analyze the memory organization required by the FFT and the two types of NTT.
The FFT inputs are 512 complex numbers stored as 1024 elements of 32 bits each.
The ML-KEM NTT inputs are 256 integers stored as 256 elements of 12 bits each. The ML-DSA inputs are 256 integers stored as 256 elements of 23 bits each. Therefore, ML-DSA elements can be accommodated in the same word size as FFT elements after zero padding to 32 bits, while pairs of ML-KEM elements can be accommodated in the same word size as single FFT elements after zero padding to pairs of 16 bits. Similar approach can also be followed for storing the FFT and NTT twiddle factors. Again, no additional memory is required to support NTT beyond the requirements of FFT.

The top-level architecture of the unified accelerator is shown in Fig. \ref{fig:top_arch}. 
Apart from the unified butterfly unit, it contains a data RAM bank, a twiddle factor ROM and control circuitry. The butterfly inputs and outputs follow an in-place data flow. The data RAM bank is implemented as a set of eight 256 $\times$ 16-bit true-dual-port memories. The twiddle factor ROM is implemented as a 1024 $\times$ 32-bit dual-port read-only memory. The unified butterfly unit is implemented with a 9-stage internal pipeline for improved performance.
It consists of four instances of the 32-bit unsigned Karatsuba multiplier described earlier.
For signed multiplication in FFT, we include 32-bit and 64-bit two's complement converter circuits respectively before and after each 32-bit unsigned multiplier.
Overall, our unified accelerator can execute either 1 FFT butterfly or 2 parallel ML-DSA NTT butterflies or 4 parallel ML-KEM NTT butterflies.

Detailed architecture of the accelerator with internal circuitry of the unified butterfly is shown in Fig. \ref{fig:detailed_arch}.
It has three operating modes (indicated by a 2-bit external input): 512-point FFT, NTT for ML-KEM (CRYSTALS-Kyber) and NTT for ML-DSA (CRYSTALS-Dilithium). The 9-stage-pipelined unified butterfly arithmetic unit has twelve 16-bit inputs and eight 16-bit outputs.
The mod\_kyber\_mul and mod\_dilithium\_mul units are used to perform Barrett reduction after multiplication in the NTT modes. The mod\_kyber\_add/sub and mod\_dilithium\_add/sub are used to reduce the addition/subtraction outputs in the NTT modes.
Clearly, apart from modular reduction circuits and multiplexors, all other circuitry in the unified FFT + NTT butterfly unit are those already required for FFT, thus highlighting efficient hardware resource sharing in our proposed architecture.

\section{Implementation Results}
\label{sec:implementation}

\begin{figure}[!t]
\centering
\includegraphics[width=3.4in]{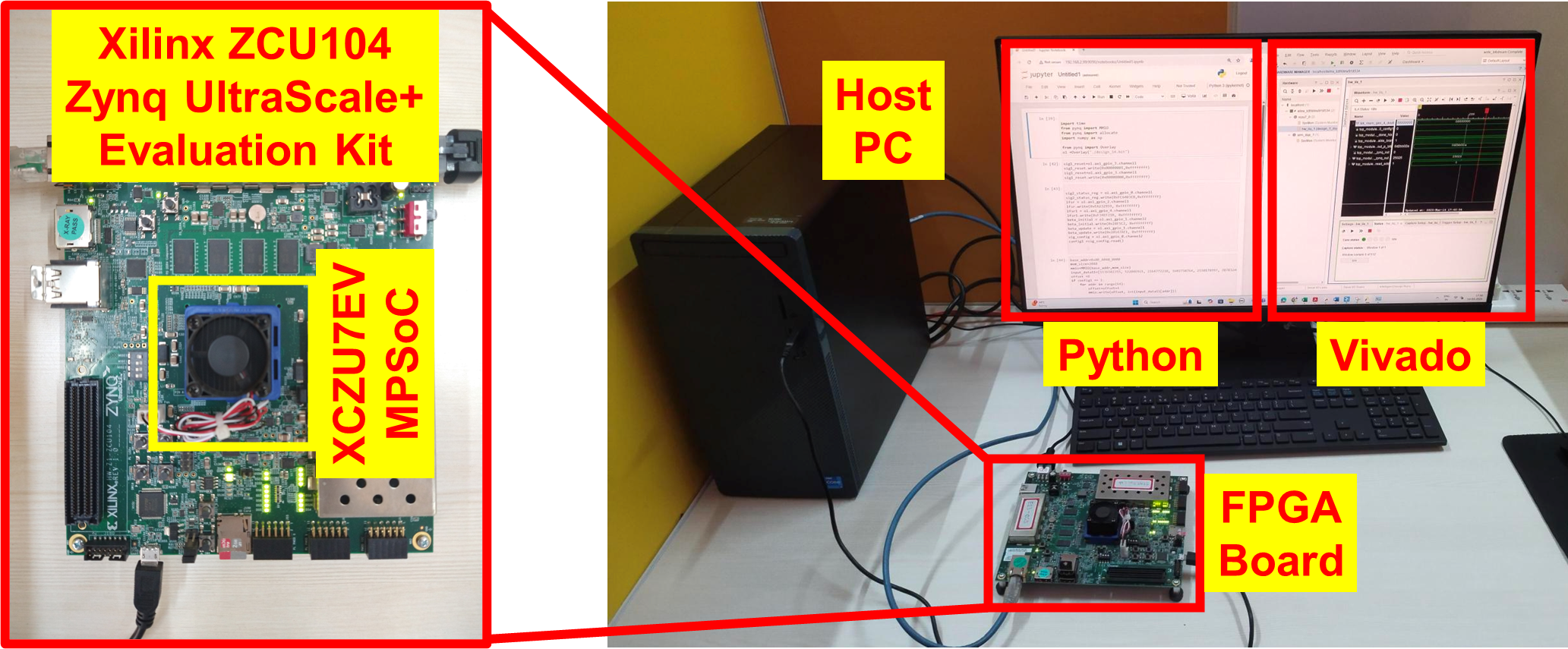}
\caption{Experimental validation setup with AMD Xilinx Zynq UltraScale+ ZCU104 MPSoC FPGA board and PYNQ interface.}
\label{fig:fpga_demo_setup}
\end{figure}

\begin{table}[!t]
\centering
\renewcommand{\arraystretch}{1.1}
\caption{Accelerator Implementation Results from ZCU104 \\ (Zynq UltraScale+ at Operating Frequency 400 MHz)}
\label{table:results_zcu104}
\setlength{\tabcolsep}{4.0pt}
\centering
\begin{tabular}{|c|c|c|c|c|c|c|c|c|c|}
\hline
\multirow{2}{*}{\textbf{{Design}}} & \multirow{2}{*}{\textbf{{Mode}}} & \multirow{2}{*}{\textbf{{LUT / FF / DSP / BRAM}}} & \textbf{{Cycle}} & \textbf{{Latency}} \\
&  &  & \textbf{{Count}} & \textbf{{($\mu$s)}} \\
\hline
\hline
{I} & ML-KEM & 3716 / 2400 / 12 / 5 & 322 & {0.80}\\
\hline
{I} & ML-DSA & 3716 / 2400 / 12 / 5 & 624 & {1.56}\\
\hline
{I} & FFT & 3716 / 2400 / 12 / 5 & 2430 & {6.08}\\
\hline
{II} & FFT & 2300 / 1898 / 12 / 5 & 2430 & {6.08}\\
\hline
{III} & ML-DSA & 2287 / 1924 / 6 / 5 & 624 & {1.56}\\
\hline
{IV} & ML-KEM & 1973 / 1246 / 4 / 4.5 & 322 & {0.80}\\
\hline
{V} & ML-KEM & 3013 / 1828 / 6 / 5 & 322 & {0.80}\\
\hline
{V} & ML-DSA & 3013 / 1828 / 6 / 5 & 624 & {1.56}\\
\hline
\end{tabular}
\end{table}

The proposed unified accelerator is implemented on a Xilinx Zynq UltraScale+ MPSoC ZCU104 Evaluation Board with an XCZU7EV-2FFVC1156E device \cite{xilinx_ultrascale} and our experimental setup is shown in Fig. \ref{fig:fpga_demo_setup}.
Verilog HDL (Hardware Description Language) is used to design the hardware accelerator, and Xilinx Vivado Design Suite version ML 2022.2 is utilized for FPGA synthesis, implementation and simulation.
Our experimental validation framework interfaces the proposed accelerator with an additional bank of ROMs populated with test vectors corresponding to expected outputs at the end of all stages of the FFT or NTT computations. These test vectors are generated using a Python script based on the twiddle factors and random input sequences / polynomials. The validation framework also includes additional test logic to compare the data stored in the data RAM bank row-wise with the golden data stored in the ROMs.
Five variants of the accelerator design are implemented: (I) FFT + ML-DSA NTT + ML-KEM NTT, (II) baseline FFT, (III) baseline ML-DSA, (IV) baseline ML-KEM, and (V) ML-DSA + ML-KEM.
These are used to study the trade-offs associated with different levels of functional unification in the same underlying architecture.
FPGA implementation results (resource utilization and performance metrics) for all five variants are shown in Table \ref{table:results_zcu104}.

\textcolor{black}{Design II (our baseline FFT) follows the architectural blueprint of the Xilinx FFT IP \cite{xilinx_fft}. Their primary difference is that Design II stores the inputs / outputs (4 KB) in 4 BRAMs compared to 2 BRAMs in \cite{xilinx_fft}, allowing multiple simultaneous reads / writes per cycle. 
We leverage this flexibility to get better performance metrics for our FFT baseline, as well as better architectural consistency with our NTT baselines, enabling fair comparison across design variants at the cost of higher BRAM usage.
Compared to Design II, Design I (FFT + ML-DSA NTT + ML-KEM NTT)} requires $\approx 62$\% more LUTs and $\approx 26$\% more FFs. The number of DSPs and BRAMs remains unchanged.
It is noteworthy that our careful design of the pipeline stages in the unified butterfly unit ensures that there is no reduction in the operating frequency (400~MHz for all the variants) due to the additional modular reduction circuitry.
This shows the trade-off between additional NTT functionality and resource utilization.
Compared to \textcolor{black}{Design V (ML-KEM + ML-DSA), Design I} requires 6 additional DSPs to accommodate two more 32-bit Karatsuba multipliers for FFT along with $\approx 23$\% more LUTs and $\approx 31$\% more FFs.

Table \ref{table:comparison} compares our unified FFT + ML-KEM / ML-DSA NTT accelerator (\textcolor{black}{Design I}) with state-of-the-art FPGA-based NTT hardware accelerators for ML-KEM and/or ML-DSA.
Despite supporting additional functionality, our implementation clearly achieves performance comparable to prior work in similar UltraScale+ FPGA platforms.
The overheads in terms of LUTs, FFs, DSPs and BRAMs are primarily due to the additional circuitry for 512-point complex FFT which is not supported by any of the previous implementations.
In particular, our unified accelerator requires 12 DSPs in order to support the fixed-point complex arithmetic for FFT and 5 BRAMs in order to store the complex inputs / outputs and twiddle factors for FFT.
Our implementation demonstrates the efficiency and flexibility of the proposed unified FFT + NTT architecture which allows the same hardware to be used for both digital signal processing as well as post-quantum lattice-based cryptography applications.

\begin{table*}[!t]
\renewcommand{\arraystretch}{1.25}
\caption{Comparison with State-of-the-Art FPGA-Based FFT, ML-KEM NTT and ML-DSA NTT Hardware Implementations}
\label{table:comparison}
\setlength{\tabcolsep}{4.0pt}
\centering
\begin{tabular}{|c|c|c|c|c|c|c|c|c|c|c|}
\hline
\multirow{2}{*}{\textbf{Design}} & \multirow{2}{*}{\textbf{Supported Algorithms}} & \textbf{FPGA} & \textbf{FPGA Resource Utilization} & \multirow{2}{*}{\textbf{Mode}} & \textbf{Freq.} & \textbf{Cycle} & \textbf{Latency} \\
& \textbf{} & \textbf{Platform} & \textbf{LUT / FF / DSP / BRAM} & & \textbf{(MHz)} & \textbf{Count} & \textbf{($\mu$s)} \\
\hline
\hline
{ISCAS '21 \cite{zhang_iscas_2021}} & ML-KEM NTT only & Artix-7 & 609 / 640 / 2 / 2 & ML-KEM & 257 & 490 & {1.9}\\
\hline
{ARITH '21 \cite{azarderakhsh_arith_2021}} & ML-KEM NTT only & Artix-7 & 801 / 717 / 4 / 2 & ML-KEM & 222 & 324 & {1.46}\\
\hline
{CARDIS '22 \cite{land_cardis_2022}} & ML-DSA NTT only & Artix-7 & 524 / 759 / 17 / 1 & ML-DSA & 311 & 533 & 1.71\\
\hline
{TCAS-I '23 \cite{gupta_tcas1_2023}} & ML-DSA NTT only & Zynq UltraScale+ & 2759 / 2037 / 4 / 7 & ML-DSA & 391 & $-$ & $-$\\
\hline
{\multirow{2}{*}{TCAS-I '23 \cite{aikata_tcas1_2023}}} & \multirow{2}{*}{ML-KEM / ML-DSA NTT} & \multirow{2}{*}{Zynq UltraScale+} & \multirow{2}{*}{3487 / 1918 / 4 / 1} & ML-KEM & \multirow{2}{*}{270} & $-$ & $-$\\
\cline{5-5}\cline{7-8}
 & & &  & ML-DSA &  & $-$ & $-$\\
\hline
{\multirow{2}{*}{TVLSI '24 \cite{li_tvlsi_2024}}} & ML-KEM NTT only & \multirow{2}{*}{Kintex UltraScale} & 1914 / 2249 / 3 / 3 & ML-KEM & 275 & $-$ & $-$\\
\cline{2-2}\cline{4-8}
 & ML-DSA NTT only &  & 5478 / 4955 / 12 / 6 & ML-DSA & 250 & 284 & 1.14\\
\hline
{\multirow{2}{*}{VLSID '24 \cite{mandal_vlsid_2024}}} & \multirow{2}{*}{ML-KEM / ML-DSA NTT} & \multirow{2}{*}{Zynq UltraScale+} & \multirow{2}{*}{2893 / 2356 / 4 / 4.5} & ML-KEM & \multirow{2}{*}{342} & 224 & 0.65\\
\cline{5-5}\cline{7-8}
 & & &  & ML-DSA &  & 512 & 1.50\\
\hline
\textcolor{black}{{Xilinx FFT IP \cite{xilinx_fft}}} & \textcolor{black}{512-point Complex FFT only} & \textcolor{black}{Zynq UltraScale+} & \textcolor{black}{1462 / 2269 / 12 / 3} & \textcolor{black}{FFT} & \textcolor{black}{456} & \textcolor{black}{3548} & \textcolor{black}{{7.78}}\\
\hline
\multirow{3}{*}{\textbf{\shortstack{This Work \\ (Design I)}}} & \multirow{3}{*}{\textbf{\shortstack{512-point Complex FFT + \\ ML-KEM / ML-DSA NTT}}} & \multirow{3}{*}{\textbf{Zynq UltraScale+}} & \multirow{3}{*}{\textbf{3716 / 2400 / 12 / 5}} & \textbf{ML-KEM} & \multirow{3}{*}{\textbf{400}} & \textbf{322} & \textbf{0.80}\\
\cline{5-5}\cline{7-8}
 &  &  &  & \textbf{ML-DSA} &  & \textbf{624} & \textbf{1.56}\\
\cline{5-5}\cline{7-8}
 &  &  &  & \textbf{FFT} &  & \textbf{2430} & \textbf{6.08}\\
\hline
\end{tabular}
\end{table*}

\section{Conclusions and Future Work}
\label{sec:conclusion}

The Number Theoretic Transform (NTT) is an indispensable tool for computing efficient polynomial multiplications in post-quantum lattice-based cryptography. It has strong resemblance with the Fast Fourier Transform (FFT) which is the most widely used algorithm for frequency-domain analysis of sequences in digital signal processing.
Following this observation, we demonstrate a unified hardware accelerator supporting both 512-point complex FFT as well as 256-point NTT for NIST post-quantum cryptography standards ML-KEM and ML-DSA.
Our proposed architecture effectively utilizes the arithmetic circuitry and memory organization required for FFT and re-purposes them for ML-KEM and ML-DSA NTT computations.
Our FPGA-based implementations achieve performance comparable to state-of-the-art ML-KEM / ML-DSA NTT accelerators, thus demonstrating the efficiency of our design.
Extension of our proposed architecture to support NTT for emerging applications of lattice-based cryptography such as homomorphic encryption \cite{acar_survey_2018} as well as efficient ASIC implementations will be explored in future work.

\section*{Acknowledgment}

This work was supported in part by a seed grant from the Indian Institute of Science, in part by the Centre of Excellence in Cyber Security (CySecK), Government of Karnataka and in part by the  Ministry of Education, Government of India.
The authors would like to thank the anonymous reviewers for their valuable feedback and Dr. Shantharam Kalipatnapu for helping with the FPGA setup.


\bibliographystyle{IEEEtran}
\bibliography{references}

\begin{thebibliography}{10}
\providecommand{\url}[1]{#1}
\csname url@samestyle\endcsname
\providecommand{\newblock}{\relax}
\providecommand{\bibinfo}[2]{#2}
\providecommand{\BIBentrySTDinterwordspacing}{\spaceskip=0pt\relax}
\providecommand{\BIBentryALTinterwordstretchfactor}{4}
\providecommand{\BIBentryALTinterwordspacing}{\spaceskip=\fontdimen2\font plus
\BIBentryALTinterwordstretchfactor\fontdimen3\font minus \fontdimen4\font\relax}
\providecommand{\BIBforeignlanguage}[2]{{%
\expandafter\ifx\csname l@#1\endcsname\relax
\typeout{** WARNING: IEEEtran.bst: No hyphenation pattern has been}%
\typeout{** loaded for the language `#1'. Using the pattern for}%
\typeout{** the default language instead.}%
\else
\language=\csname l@#1\endcsname
\fi
#2}}
\providecommand{\BIBdecl}{\relax}
\BIBdecl

\bibitem{shor_quantum_1997}
P.~W. {Shor}, ``{Polynomial-Time Algorithms for Prime Factorization and Discrete Logarithms on a Quantum Computer},'' \emph{SIAM Journal of Computing}, vol.~26, no.~5, pp. 1484--1509, Oct. 1997.

\bibitem{nist_pqc0_2016}
NIST, ``{Report on Post-Quantum Cryptography},'' IR 8105, 2016.

\bibitem{nist_pqc1_2019}
------, ``{Status Report on the First Round of the NIST Post-Quantum Cryptography Standardization Process},'' IR 8240, 2019.

\bibitem{nist_pqc2_2020}
------, ``{Status Report on the Second Round of the NIST Post-Quantum Cryptography Standardization Process},'' IR 8309, 2020.

\bibitem{nist_pqc3_2020}
------, ``{Status Report on the Third Round of the NIST Post-Quantum Cryptography Standardization Process},'' IR 8413, 2022.

\bibitem{crystals_kyber_2021}
R.~{Avanzi} \emph{et~al.}, ``{CRYSTALS-Kyber -- Algorithm Specifications and Supporting Documentation},'' NIST, Tech. Rep., 2021.

\bibitem{crystals_dilithium_2021}
S.~{Bai} \emph{et~al.}, ``{CRYSTALS-Dilithium -- Algorithm Specifications and Supporting Documentation},'' NIST, Tech. Rep., 2021.

\bibitem{nist_mlkem_2023}
{NIST}, ``{Module-Lattice-Based Key-Encapsulation Mechanism Standard},'' FIPS 203, Aug. 2024.

\bibitem{nist_mldsa_2023}
------, ``{Module-Lattice-Based Digital Signature Standard},'' FIPS 204, Aug. 2024.

\bibitem{peikert_decade_2016}
C.~{Peikert}, ``{A Decade of Lattice Cryptography},'' \emph{Foundations and Trends in Theoretical Computer Science}, vol.~10, no.~4, pp. 283--424, Mar. 2016.

\bibitem{cormen_algo_2009}
T.~H. {Cormen}, C.~E. {Leiserson}, R.~L. {Rivest}, and C.~{Stein}, \emph{{Introduction to Algorithms}}, 3rd~ed.\hskip 1em plus 0.5em minus 0.4em\relax The MIT Press, 2009.

\bibitem{banerjee_sapphire_2019}
U.~{Banerjee}, T.~S. {Ukyab}, and A.~P. {Chandrakasan}, ``{Sapphire: A Configurable Crypto-Processor for Post-Quantum Lattice-based Protocols},'' \emph{{IACR Transactions on Cryptographic Hardware and Embedded Systems}}, vol. 2019, no.~4, pp. 17--61, Aug. 2019.

\bibitem{banerjee_iscas_2020}
U.~{Banerjee}, S.~{Das}, and A.~P. {Chandrakasan}, ``{Accelerating Post-Quantum Cryptography using an Energy-Efficient TLS Crypto-Processor},'' in \emph{2020 IEEE International Symposium on Circuits and Systems (ISCAS)}, Oct. 2020, pp. 1--5.

\bibitem{bos_usenix_2022}
J.~W. Bos, J.~Renes, and C.~van Vredendaal, ``{Post-Quantum Cryptography with Contemporary Co-Processors: Beyond Kronecker, Sch{\"o}nhage-Strassen \& Nussbaumer},'' in \emph{2022 USENIX Security Symposium}, 2022, pp. 3683--3697.

\bibitem{zhang_iscas_2021}
C.~Zhang, D.~Liu, X.~Liu, X.~Zou, G.~Niu, B.~Liu, and Q.~Jiang, ``{Towards Efficient Hardware Implementation of NTT for Kyber on FPGAs},'' in \emph{2021 IEEE International Symposium on Circuits and Systems (ISCAS)}, 2021, pp. 1--5.

\bibitem{azarderakhsh_arith_2021}
M.~Bisheh-Niasar, R.~Azarderakhsh, and M.~Mozaffari-Kermani, ``{High-Speed NTT-based Polynomial Multiplication Accelerator for Post-Quantum Cryptography},'' in \emph{2021 IEEE Symposium on Computer Arithmetic (ARITH)}, 2021, pp. 94--101.

\bibitem{land_cardis_2022}
G.~Land, P.~Sasdrich, and T.~G{\"u}neysu, ``{A Hard Crystal - Implementing Dilithium on Reconfigurable Hardware},'' in \emph{International Conference on Smart Card Research and Advanced Applications (CARDIS)}, 2022, pp. 210--230.

\bibitem{gupta_tcas1_2023}
N.~Gupta, A.~Jati, A.~Chattopadhyay, and G.~Jha, ``{Lightweight Hardware Accelerator for Post-Quantum Digital Signature CRYSTALS-Dilithium},'' \emph{IEEE Transactions on Circuits and Systems I: Regular Papers}, vol.~70, no.~8, pp. 3234--3243, 2023.

\bibitem{aikata_tcas1_2023}
A.~Aikata, A.~C. Mert, M.~Imran, S.~Pagliarini, and S.~S. Roy, ``{KaLi: A Crystal for Post-Quantum Security Using Kyber and Dilithium},'' \emph{IEEE Transactions on Circuits and Systems I: Regular Papers}, vol.~70, no.~2, pp. 747--758, 2023.

\bibitem{li_tvlsi_2024}
B.~Li, Y.~Yan, Y.~Wei, and H.~Han, ``{Scalable and Parallel Optimization of the Number Theoretic Transform Based on FPGA},'' \emph{IEEE Transactions on Very Large Scale Integration (VLSI) Systems}, vol.~32, no.~2, pp. 291--304, 2024.

\bibitem{mandal_vlsid_2024}
S.~Mandal and D.~B. Roy, ``{KiD: A Hardware Design Framework Targeting Unified NTT Multiplication for CRYSTALS-Kyber and CRYSTALS-Dilithium on FPGA},'' in \emph{International Conference on VLSI Design (VLSID)}, 2024, pp. 455--460.

\bibitem{thanawala_fft_2023}
N.~Thanawala, H.~Nejatollahi, and N.~Dutt, ``{Accelerating Polynomial Multiplication for RLWE using Pipelined FFT},'' Cryptology ePrint Archive, Report 2023/1815, 2023.

\bibitem{karatsuba_mul_1962}
A.~{Karatsuba} and Y.~{Ofman}, ``{Multiplication of Many-Digital Numbers by Automatic Computers},'' \emph{Proceedings of USSR Academy of Sciences}, vol. 145, no.~7, pp. 293--294, 1962.

\bibitem{barrett_red_1986}
P.~{Barrett}, ``{Implementing the Rivest Shamir and Adleman Public Key Encryption Algorithm on a Standard Digital Signal Processor},'' in \emph{Advances in Cryptology (CRYPTO '86')}, Aug. 1986, pp. 311--323.

\bibitem{xilinx_ultrascale}
{Xilinx Inc.}, ``{UltraScale Architecture: Staying a Generation Ahead with an Extra Node of Value},'' \url{https://www.xilinx.com/products/technology/ultrascale.html}.

\bibitem{xilinx_fft}
------, ``{LogiCORE IP - Fast Fourier Transform v9.1},'' \url{https://www.xilinx.com/products/intellectual-property/fft.html}.

\bibitem{acar_survey_2018}
A.~Acar, H.~Aksu, A.~S. Uluagac, and M.~Conti, ``{A Survey on Homomorphic Encryption Schemes: Theory and Implementation},'' \emph{ACM Computing Surveys}, vol.~51, no.~4, pp. 1--35, 2018.

\end{thebibliography}

\end{document}